# A Rule-Based Relational XML Access Control Model in the Presence of Authorization Conflicts

Ali Alwehaibi
Dept. of Computational Science
NC A&T State University
Greensboro, NC 27405 USA
asalweha@aggies.ncat.edu

Mustafa Atay*
Dept. of Computer Science
Winston-Salem State University
Winston-Salem, NC 27110 USA
ataymu@wssu.edu

*Abstract*—There is considerable amount of sensitive XML data stored in relational databases. It is a challenge to enforce node level fine-grained authorization policies for XML data stored in relational databases which typically support table and column level access control. Moreover, it is common to have conflicting authorization policies over the hierarchical nested structure of XML data. There are a couple of XML access control models for relational XML databases proposed in the literature. However, to our best knowledge, none of them discussed handling authorization conflicts with conditions in the domain of relational XML databases. Therefore, we believe that there is a need to define and incorporate effective fine-grained XML authorization models with conflict handling mechanisms in the presence of conditions into relational XML databases. We address this issue in this study.

## I. INTRODUCTION

XML documents over the Web and across corporate networks include critical governmental, financial, medical and scientific information with sensitive data. It is crucial to protect the sensitive data from the access of unauthorized users. Enforcing access restrictions on XML data has become critical in order to have efficient mechanisms to securely store and query XML.

In general, access restriction mechanisms are defined by a set of authorization policies which typically grant or deny access to objects of a data set for specific users. An authorization policy set is likely to include conflicting policies such as grant and deny, grant and partial deny (conditional deny), partial grant (conditional grant) and partial deny, defined on the same object. These authorization conflicts need to be recognized and handled effectively in a secure and reliable access restriction mechanism.

Several researchers have proposed to use the mature relational database technology to store, secure and query XML data [1, 5, 6, 11, 12]. Although relational databases support table level and tuple level access control mechanism, they do not provide node-level restriction mechanisms for the hierarchical XML data. Moreover, a relational XML storage needs to be equipped with a fine-grained conflict resolution support.

We propose a rule-based access control model for relational XML databases in this paper. In the proposed access control model, XML authorization rules are converted into relational tuples to be stored in a relational table. Overall security check is conducted inside the relational database. The authorization conflicts are dealt with effectively using a centralized authorization table in the database.

The main contributions of this paper include the followings:

i.   We augment our test bed relational XML Database (XML2REL) [1] with an access control model which enables document authors to define and enforce XML authorization rules which include conflicting grant and deny privileges over XML elements and subtrees stored in a relational database. The XAR2RAR algorithm proposed in this paper takes an XML authorization document and translates it into a relational authorization table to be used in secure query processing.

ii.  We define the concepts of *absolute conflicts* and *partial conflicts* to deal with conflicting authorization policies in a fine-grained access control model.

## II. MOTIVATION

The authorization conflicts occur in various cases. The conflicts in defining authorization policies may arise due to different reasons. If the authorization policies are not issued by a central division in an enterprise, it is likely to have redundant or conflicting policies proposed by different departments. Even if the authorization policies are issued by a central authority, these policies will not be static. Existing policies may need to be modified as the business rules change or the enterprise evolves. Hence, the existing security rules and the newly proposed rules may cause authorization conflicts. In either case, those conflicts need to be resolved correctly and properly.

A fine-grained access control model:

i.   should allow authorized users to access every single part of the data set that they are permitted.

ii. should prevent authorized users from accessing any part of the data set that they are not permitted.

In a fine grained access control model, an effective and correct conflict resolution varies from case to case. Therefore, identifying various conflicting situations and dealing with them accordingly is a challenging important task in a fine-grained access control model. The following list includes examples of various conflicting authorization cases based on the security rules defined for accessing student GPA's (Grade Point Average) in an educational domain:

1. When having grant and deny authorizations on the same object with only grant having a condition such as "staff is granted access to GPA > 2.0" and "staff is denied access to GPA".

2. When having grant and deny authorizations on the same object with only deny having a condition such as "staff is granted access to GPA" and "staff is denied access to GPA < 2.0".

3. When having grant and deny authorizations on the same object with both having conditions and the condition of grant is the subset of the condition of deny such as "staff is granted access to GPA > 3.0" and "staff is denied access to GPA > 2.0".

4. When having grant and deny authorizations on the same object with both having conditions and the condition of deny is the subset of the condition of grant such as "staff is granted access to GPA < 3.0" and "staff is denied access to GPA < 2.0".

A fine-grained access control model deals with each one of the various conflict cases differently. For example, while our proposed fine-grained access control model eliminates the grant rule in the first case, in the second case, it does not eliminate but revise the grant rule with the reverse of the deny condition as follows: "staff is granted access to GPA >= 2.0". In case #4, grant rule is not eliminated but revised with the difference of ranges of grant and deny conditions as follows: "staff is granted access to 2.0 <= GPA < 3.0"

Each case needs to be handled properly and differently so that the access control system is ensured to generate and retain fine-grained authorization policies which is a challenging task. We believe that there is need to define and incorporate fine-grained XML authorization models with effective conflict handling mechanisms in the presence of conditional security policies into relational XML databases. We address this issue in this study.

## III. RELATED WORK

Although a number of access control models were proposed for native XML databases in the literature [2, 3, 4, 7, 8, 9, 10, 14], there are lesser access control models proposed by researchers for relational XML databases [5, 6, 8, 11, 12]. In [5, 6], authors proposed schema-oblivious XML access control techniques using relational databases. In [11, 12], authors proposed schema-based XML access control approaches for relational databases. However, handling authorization conflicts in the presence of conditions is not elaborated in these studies.

The authorization policies in XML security models are mainly defined either as external subset or internal subset [5]. In external subset approach, the policies are defined in an external file based on the content and structure of the underlying XML document which is denoted by a DTD or XML Schema. Internal subset approach defines and inlines additional tags within the target XML document to incorporate the security information. Internal subset overrides the external subset if they occur together. We choose external subset approach in our study.

Various conflict resolution policies proposed in the literature solve conflicts with general policies such as "denials take precedence or deny overrides" [3, 8]. However, these policies do not consider the cases when the subject is conditionally denied accessing part of the object that was previously granted access. We propose a fine-grained access control model for relational XML databases which address all aspects of handling authorization conflicts including conditional ones in this paper.

## IV. PROPOSED AUTHORIZATION MODEL

In this section, we propose a fine-grained authorization model for relational XML databases. We adopted our previously defined deny-only access control model [11] and enhanced it with the inclusion of grant rules along with deny rules and conflict resolution policies. We also changed the default semantics from *grant* to *deny* in this enhanced authorization model.

In our authorization model, an access control policy contains a set of authorization rules. In this access control model, an authorization rule is a tuple of the form *(subject, object, condition, action, type, mode)* where

- *subject* denotes to a group of users
- *object* refers to the group of data that the subject is concerned with
- *condition* denotes the optional predicate applied to the *object*
- *action* refers to the type of action (select, update, delete, etc.) that the subject is denied or allowed
- *type* indicates whether the rule affects only the object or propagates to the descendants of the object
- *mode* indicates whether the *action* is grant or deny

In our enhanced access control model, we consider both *grant* and *deny* rules. We introduced the *mode* parameter to enable defining opposite authorization rules with *grant* and *deny* modes unlike our previous model where only deny rules are allowed. We introduce conditional authorization rules with the use of *condition* parameter in authorization rules. Therefore, various authorization conflicts may frequently occur in the proposed access control model.

We only consider *select* action for simplicity in this paper. The default semantic of the proposed access control model is *deny*. Therefore, no one can access any part of the XML document unless there is a rule which allows a user to access a specific part, element or attribute of the XML document.

Authorization conflicts occur if two authorization rules with opposite operation modes, namely grant and deny, are defined on the same object. We categorize authorization conflicts as *absolute conflicts* and *partial conflicts* and deal with them accordingly. We introduce the notions of absolute conflicts and partial conflicts in the following definitions.

*Definition 4.1 (Absolute Authorization Conflict).* An authorization rule can be denoted as $Rule_{sbj}(Obj[predicate])$ where Rule indicates Grant or Deny, sbj is the subject of the Rule, Obj is the object concerned by the Rule and predicate is the condition defined for the Rule. If the rules $Grant_s(O[predicate_g])$ and $Deny_s(O[predicate_d])$ are given for the same subject S, object O and if $predicate_g \subseteq predicate_d$, then, Absolute Conflict is said to occur between these rules.

*Definition 4.2 (Partial Authorization Conflict).* An authorization rule can be denoted as $Rule_{sbj}(Obj[predicate])$ where Rule indicates Grant or Deny, sbj is the subject of the Rule, Obj is the object concerned by the Rule and predicate is the condition defined for the Rule. If the rules $Grant_s(O[predicate_d])$ and $Deny_s(O[predicate_d])$ are given for the same subject S, object O and if one of the following conditions holds: i) $predicate_d \subset predicate_g$. ii) $predicate_d \cap predicate_g \neq \varnothing$ AND $predicate_g - predicate_d \neq \varnothing$ AND $predicate_d - predicate_g \neq \varnothing$ then, Partial Conflict is said to occur between these rules.

Our general conflict resolution policy is "*the latter rule overrides*" in the presence of *absolute conflicts*. In addition, we fine-tune our conflict resolution policy with the consideration of *partial conflicts* which occur due to the predicates. If a deny rule conflicts with a grant rule and if at least one of them is defined with a predicate, our algorithm analyzes the predicate(s) and partially override the former rule if there exists a *partial conflict*. Therefore, in the case of partial conflicts, conflict resolution policy becomes "*the latter rule partially overrides*".

We consider simple XPath predicates to test target text nodes or attributes within XPath expressions of the rule tuples. We do not consider twig pattern matching and predicates incurred by twig pattern matching in this study. We use a subset of XPath axes, namely *child*, *descendant* and *attribute* axes, to identify the objects in a rule tuple.

The propagation of an authorization rule to the descendants of the XML object node is not enforced automatically. The *type* parameter in a rule tuple determines the scope of an authorization rule. While a local rule only affects the object node itself, a recursive rule is applied to the object node as well as the descendants of the object node.

A sample XML authorization rule is shown in Figure 1. The *condition* parameter of a rule tuple is combined with the XPath expression of the object *parameter*. The rule tuple *(staff, //zip, [.>48000], select, L, grant)* indicates that any *zip* element which is greater than 48000 regardless of its ancestors is granted *select* access for *staff*. The *type L* denotes that this rule is local and does not affect the descendants of *zip* node if there is any descendant node.

```
<rule>
    <subject>staff</subject>
    <object>//zip[.>48000]</object>
    <action>select</action>
    <type>L</type>
    <mode>grant</mode>
</rule>
```

Figure 1. Sample authorization policy rule

## V. XML-TO-RELATIONAL SECURITY POLICY CONVERSION

The Security Policy Conversion module translates policy rules and users' information from XML to relational tables. The XML document for users information may include *userID, password, first* and *last name* and *role* of each user. It is straightforward to translate XML subtrees of users' information to relational rows in a table.

XML policy subtrees are mapped into relational tuples to be stored in XML Authorization Table (XAT) in the target relational database. The XAT table is the central place of authorization control in the proposed access control model. The XAT table stores only the grant rules since the default semantic of the proposed access control model is *deny*. Thus, when a query is given against the underlying database, the security system allows the query if there is a policy rule in the XAT table granting the user to execute the requested query against the specified object. Deny policy rules are not stored in the XAT table but dealt with accordingly.

During the mapping process of XML Authorization Rules (XAR) to Relational Authorization Rules (RAR), if the *mode* parameter of the XML authorization rule is *grant*, then a relational rule tuple is inserted into the XAT table. A relational authorization rule (RAR) is deleted from the XAT table if there is a corresponding XAR rule denoted with *deny* mode in the XML authorization document which causes an *absolute conflict*. If an XAR deny rule partially conflicts with a RAR grant rule in the XAT table, then, the RAR rule is updated with a relevant predicate.

We do not map *mode* and *type* parameters of an XAR rule to a RAR rule tuple. The *mode* parameter is not mapped since all the RAR rules in XAT table are grant rules. The *type* parameter is not mapped to a RAR rule tuple since all RAR rules in XAT table are local as the recursive rules are expanded to local ones during the mapping process.

It is a challenge to translate an XML policy subtree to one or more relational tuples in XML Authorization Table (XAT). While it is straightforward to map *subject* and *action* parameters of an XML policy rule to relational columns in the XAT table, there are several issues included in translating an

*object* parameter along with its *condition* parameter to its relational equivalents. The issues involved in translating *object* parameters of XAR rules to RAR rules are elaborated in [11]. These issues are primarily due to the '//' axis, recursive scope of the policy and manipulation of the conditions. Conditions can get complex with multiple logical operators. In addition, conditions can be introduced to other nodes besides the context node which increases the complexity of mapping.

```
00 Algorithm XAR2RAR
01 Input: XAR_Rules Authorization.xml
02 Output: RAR_Rules XAT
03 Begin
04 For each Rule_i in Authorization.xml
05   Predicate, newPredicate = null;
06   If (Object contains Predicate) then Predicate=Object.Predicate End If
07   If (Type='L') then
08     If (Object contains '//') then
09       ObjectPattern = Object.replace ("//", "%")
10       ObjectSet = "Select path From AllPaths Where path like $ObjectPattern"
11     Else
12       ObjectSet = Object
13     End If
14     For each object_i in ObjectSet
15       If @Mode='Grant' then
16         Insert into XAT (Subject, object_i, Predicate, Action)
17       Else  /* Mode = 'Deny' */
18         If there exists absolute conflict then
19 Delete From XAT Where (XAT.Subject=Subject AND XAT.Object=object_i AND
                          XAT.Action=Action)
20         Else If there exists partial conflict then
21 Update XAT Set XAT.Predicate=newPredicate Where (XAT.Subject=Subject
                AND XAT.Object=object_i AND XAT.Action=Action)
22         End If
23       End If
24     End If
25   End For
26   Else  /* Type = 'R' */
27     If (Object contains '//') then
28       ObjectPattern = Object.replace ("//", "%")
29       ObjectSet = "Select Path From AllPaths
                     Where (path like $ObjectPattern OR path like $ObjectPattern/%)"
30     Else
31       ObjectSet = "Select path From AllPaths
                     Where (path like $Object OR path like $Object/%)"
32     End If
33     For each object_i in objectSet
34       If @Mode='Grant' then
35         Insert into XAT (Subject, object_i, Predicate, Action)
36       Else  /* Mode = 'Deny' */
37         If there exists absolute conflict then
38 Delete from XAT Where (XAT.Subject=Subject AND XAT.Object=object_i
                          AND XAT.Action=Action)
39         Else
40           If there exists partial conflict then
41 Update XAT Set XAT.Predicate=newPredicate Where (XAT.Subject=Subject
                AND XAT.Object=object_i AND XAT.Action=Action)
42           End If
43         End If
44       End If
45     End For
46   End If
47 End For
48 End
```

Figure 2. XAR2RAR Algorithm

We only consider simple conditions for the context node of a path expression for an *object*. The *condition* parameter plays an important role in detecting various conflicts. There are several issues involved in manipulating *condition* parameters

(predicates) during the XAR-to-RAR mapping process. These issues include the followings:

- If the operation mode for the XAR rule is *grant,* condition should be extracted and inserted into the XAT table along with other XAR parameters
- If the operation mode for the XAR rule is *deny*, the type of conflict should be detected analyzing the *condition* parameters of both XAR deny rule and the corresponding RAR grant rule
- If there is a partial conflict then a modified condition (new Predicate) needs to be calculated to update the existing RAR rule tuple

We propose XAR2RAR algorithm to convert XML access control rules (XAR) into relational access control rules (RAR) as well as to deal with authorization conflicts. This algorithm detects and resolves authorization conflicts as it translates the XAR rules in the input Authorization.xml document into the RAR rules in the output XML Authorization Table (XAT). The RAR rules in the XAT table are used to enforce the access control policies for XML data stored in a RDBMS. XAR2RAR algorithm is given in Figure 2.

XAR2RAR algorithm processes each XAR rules defined in Authorization.xml within a loop. Firstly, it checks for the t*y*pe parameter. The authorization rule is applied at the node level when t*y*pe is local (L). When type is recursive (R), then the rule is applied at the node level as well as at the subtree rooted at that node. If the path expression of the *Object* has descendant axes then the XAR2RAR algorithm creates an ObjectPattern to find the set of all matching paths for that *Object.* This set of matching paths is *called ObjectSet* and obtained from a global path table (*AllPaths*) which includes all existing paths in a given XML document. We adapt the idea of introducing a global path table to our mapping scheme from XRel approach [13]. If the *type* is recursive, XAR2RAR algorithm places a path expression for the *Object* as well as the path expressions for all the descendants of the *Object* into the *ObjectSet*. Each path expression in the *ObjectSet* is processed as follows:

- If the mode of the XAR rule is *grant* then a RAR rule tuple is inserted into the XAT table (lines 16 and 35)
- If the mode of the XAR rule is *deny* and the rule causes an *absolute conflict* then the conflicting RAR rule tuple is deleted from the XAT table (lines 19 and 38)
- If the mode of the XAR rule is *deny* and the rule causes a *partial conflict* then the conflicting RAR rule tuple is updated in the XAT table (lines 21 and 41)

VI.     ILLUSTRATIVE CASE STUDY

We illustrate the proposed authorization model with a case study in this section. The *department* benchmark data set is used as our sample data set in the case study. The *department* data set's schema (DTD) is shown in Figure 3.

The first set of XAR rules given against the *department* data set are described in the XML document *Authorization1.xml* which is shown in Figure 4.

```
<!ELEMENT department (deptname, gradstudent* , staff*, faculty*, undergradstudent*)>
<!ELEMENT gradstudent (name, phone, email, address, office?, url?, gpa)>
<!ELEMENT staff (name, phone, email, office?)>
<!ELEMENT faculty (name, phone, email, office)>
<!ELEMENT undergradstudent (name, phone, email, address, gpa)>
<!ELEMENT name (lastname?,firstname)>
<!ELEMENT address (city, state, zip)>
<!ELEMENT deptname (#PCDATA)>
<!ELEMENT city (#PCDATA)>
<!ELEMENT state (#PCDATA)>
<!ELEMENT zip (#PCDATA)>
<!ELEMENT office (#PCDATA)>
<!ELEMENT phone (#PCDATA)>
<!ELEMENT lastname (#PCDATA)>
<!ELEMENT firstname (#PCDATA)>
<!ELEMENT url (#PCDATA)>
<!ELEMENT gpa (#PCDATA)>
<!ELEMENT email (#PCDATA)>
```

Figure 3. DTD of department data set

Our XAR2RAR algorithm translates XML access control rules into relational tuple(s) in XAT table and deals with authorization conflicts as it parses XML authorization documents.

```
<?xml version="1.0" encoding="utf-8"?>
<rules>
    <rule>
        <subject>staff</subject>
        <object>//gpa</object>
        <action>Select</action>
        <type>L</type>
        <mode>Grant</mode>
    </rule>
    <rule>
        <subject>staff</subject>
        <object>/department/undergradstudent/address</object>
        <action>Select</action>
        <type>R</type>
        <mode>Grant</mode>
    </rule>
    <rule>
        <subject>staff</subject>
        <object>/gradstudent/zip[.<60000]</object>
        <action>Select</action>
        <type>L</type>
        <mode>Grant</mode>
    </rule>
    <rule>
        <subject>faculty</subject>
        <object>/department/undergradstudent/address</object>
        <action>Select</action>
        <type>R</type>
        <mode>Grant</mode>
    </rule>
    <rule>
        <subject>faculty</subject>
        <object>/gradstudent/zip[.<70000]</object>
        <action>Select</action>
        <type>L</type>
        <mode>Grant</mode>
    </rule>
</rules>
```

Figure 4. Authorization1.xml document

When the first XAR rule is parsed, the object's path expression *//gpa* is extended to its corresponding path expressions. Then, two RAR tuples with the below objects are inserted into XAT table to grant *select* access to all *gpa* nodes for *staff*.

- */department/gradstudent/gpa*
- */department/undergradstudent/gpa*

When the second XAR rule is parsed, firstly, the below expanded path expressions of */department/undergradstudent/address* are extracted from *AllPaths*:

- */department/undergradstudent/address*
- */department/undergradstudent/address/city*
- */department/undergradstudent/address/state*
- */department/undergradstudent/address/zip*

Then, four RAR tuples with the above objects are inserted into XAT table which grant *select* access to all *address* subtrees of undergraduate students for *staff*.

The third XAR rule shows an example of a conditional authorization rule. The following target object path is granted access for staff subjects for zip codes less than 60000.

- */department/gradstudent/address/zip[.<60000]*

The fourth and fifth rules are processed similar to the second and third rules except they are defined for faculty subjects. The XAT table populated by the XAR2RAR algorithm for the above Authorization1.xml document is given in Table I.

TABLE I. XAT TABLE AFTER AUTHORIZATION1.XML PARSED

| XAT Table | | | |
|---|---|---|---|
| Subject | Object | Predicate | Action |
| staff | /department/gradstudent/gpa | - | Select |
| staff | /department/undergradstudent/gpa | - | Select |
| staff | /department/undergradstudent/address | - | Select |
| staff | /department/undergradstudent/address/city | - | Select |
| staff | /department/undergradstudent/address/state | - | Select |
| staff | /department/undergradstudent/address/zip | - | Select |
| staff | /department/gradstudent/address/zip | . < 60000 | Select |
| faculty | /department/undergradstudent/address | - | Select |
| faculty | /department/undergradstudent/address/city | - | Select |
| faculty | /department/undergradstudent/address/state | - | Select |
| faculty | /department/undergradstudent/address/zip | - | Select |
| faculty | /department/gradstudent/address/zip | . < 70000 | Select |

The Authorization2.xml document given in Figure 5 includes additional set of authorization rules which conflict with the ones introduced in Authorization1.xml shown in Figure 4.

First rule in Authorization2.xml denies access to subtrees of address nodes of undergraduate students for *staff*. Since we have grant rules defined for the same target objects, this raises *absolute conflicts* and all of the four tuples corresponding to those objects are deleted from the XAT table.

The second rule denies access to zip codes of graduate students if the zip code is less than 70000 for *staff*. The XAT table includes grant rule on zip code for staff subjects if the graduate student's zip code is less than 60000. Since (zip<60000) $\subseteq$ (zip<70000), this situation raises an *absolute conflict* and the corresponding tuple is deleted from the XAT table.

The third rule in Authorization2.xml denies access to GPA nodes which are less than 2.0 for *staff*. This rule conflicts with the RAR grant rule for GPA nodes. However the grant rule is unconditional and the deny rule is conditional. Since (GPA<2.0) ⊂ U, this situation raises a *partial conflict.* Thus, we do not delete the corresponding grant tuples in XAT table but update them with the new predicate (GPA >= 2.0).

```
<?xml version="1.0" encoding="utf-8"?>
<rules>
        <rule>
                <subject>staff</subject>
                <object>/undergradstudent//address</object>
                <action>Select</action>
                <type>R</type>
                <mode>Deny</mode>
        </rule>
        <rule>
                <subject>staff</subject>
                <object>/gradstudent//zip[.<70000]</object>
                <action>Select</action>
                <type>L</type>
                <mode>Deny</mode>
        </rule>
        <rule>
                <subject>staff</subject>
                <object>//gpa[.<2.0]></object>
                <action>Select</action>
                <type>L</type>
                <mode>Deny</mode>
        </rule>
        <rule>
                <subject>faculty</subject>
                <object>/gradstudent//zip[.>60000]</object>
                <action>Select</action>
                <type>L</type>
                <mode>Deny</mode>
        </rule>
</rules>
```

Figure 5. Authorization2.xml document

The last rule in Authorization2.xml denies access to zip codes of graduate students if the zip code is greater than 60000 for *faculty* subjects. The XAT table includes grant rule for *faculty* subjects if the graduate student's zip code is less than 70000. Since (zip>60000) ∩ (zip<70000) ≠∅ and differences are not empty set either, this situation raises a *partial conflict* and the corresponding RAR tuple in XAT table is updated with the revised predicate (zip <= 60000).

The XAT table populated by the XAR2RAR algorithm after the Authorization2.xml parsed is shown in Table II.

TABLE II XAT TABLE AFTER AUTHORIZATION2.XML PARSED

| XAT Table | | | |
|---|---|---|---|
| Subject | Object | Predicate | Action |
| staff | /department/gradstudent/gpa | .>= 2.0 | Select |
| staff | /department/undergradstudent/gpa | .>= 2.0 | Select |
| faculty | /department/gradstudent/address | - | Select |
| faculty | /department/gradstudent/address/city | - | Select |
| faculty | /department/gradstudent/address/state | - | Select |
| faculty | /department/gradstudent/address/zip | - | Select |
| faculty | /department/undergradstudent/address/zip | .<= 60000 | Select |

XAT table is central in the process of secure translation of XML queries into SQL queries in our relational XML database system. First, an input XML query is mapped into an SQL query. Then it is rewritten based on the security information obtained from the XAT table. The rewritten secure query returns the permitted query results.

## VII. CONCLUSIONS

We proposed a fine-grained access control model with conflict handling mechanisms in the presence of conditions for relational XML databases. We define the concepts of *absolute conflicts* and *partial conflicts* which are instrumental in evaluating and resolving authorization conflicts.

Our proposed algorithm XAR2RAR translates XML authorization rules (XAR) into relational authorization rules (RAR) and stores them in a relational table. XAR2RAR algorithm resolves conflicts due to conditions effectively and deals with overall authorization conflicts elegantly using a centralized authorization table (XAT) in the target relational database.

We introduced conflicts due to grant-deny authorization policies defined on common objects. We did not look into grant-grant or deny-deny conflicts. In case of *partial conflicts*, the underlying RAR rule needs to be updated with a relevant new predicate. The new predicate is produced differently for various situations. We consider elaborating other types of authorization conflicts and introducing a new predicate production algorithm for different cases of *partial conflicts* as potential future work.